\documentclass[twocolumn,showpacs,preprintnumbers,superscriptaddress]{revtex4}
\usepackage{graphicx}
\usepackage{dcolumn}
\usepackage{bm}
\usepackage{color}
\usepackage{amsmath}
\usepackage{amssymb}
\usepackage{amsfonts}
\usepackage{esint}
\usepackage{times}

\begin{document}

\title{Simultaneously exciting two atoms with photon-mediated Raman
interaction}
\author{Peng Zhao}
\affiliation{National Laboratory of Solid State Microstructures, \\
School of Physics, Nanjing University, Nanjing 210093, China}
\author{Xinsheng Tan}
\affiliation{National Laboratory of Solid State Microstructures, \\
School of Physics, Nanjing University, Nanjing 210093, China}
\author{Haifeng Yu}
\email{hfyu@nju.edu.cn}
\affiliation{National Laboratory of Solid State Microstructures, \\
School of Physics, Nanjing University, Nanjing 210093, China}
\affiliation{Synergetic Innovation Center of Quantum Information $\&$ Quantum Physics, \\
University of Science and Technology of China, Hefei, Anhui 230026, China}
\author{Shi-Liang Zhu}
\affiliation{National Laboratory of Solid State Microstructures, \\
School of Physics, Nanjing University, Nanjing 210093, China}
\affiliation{Synergetic Innovation Center of Quantum Information $\&$ Quantum Physics, \\
University of Science and Technology of China, Hefei, Anhui 230026, China}
\author{Yang Yu}
\affiliation{National Laboratory of Solid State Microstructures, \\
School of Physics, Nanjing University, Nanjing 210093, China}
\affiliation{Synergetic Innovation Center of Quantum Information $\&$ Quantum Physics, \\
University of Science and Technology of China, Hefei, Anhui 230026, China}
\date{\today }

\begin{abstract}
We propose an approach to simultaneously excite two atoms by using
cavity-assisted Raman process in combination with cavity photon-mediated
interaction. The system consists of a two-level atom and a $\Lambda $-type
or V-type three-level atom, which are coupled together with a cavity mode.
Having derived the effective Hamiltonian, we find that under certain
circumstances a single photon can simultaneously excite two atoms. In
addition, multiple photons and even a classical field can also simultaneously
excite two atoms. As an example, we show a scheme to realize our proposal in
a circuit QED setup, which is artificial atoms coupled with a cavity. The
dynamics and the quantum statistical properties of the process are
investigated with experimentally feasible parameters.
\end{abstract}

\maketitle


\section{Introduction}

The light matter interaction has been an important topic since the
foundation of the modern physics \cite{R1,R2}. Recent activities with the
aim for realizing quantum information process add more significance on this
area since photons provide a convenient medium to control and couple quantum
systems \cite{R3,R4,R5,R6}. Among these efforts, two-photon absorption and
emission process, where two photons are adsorbed or emitted simultaneously
by an atom or a molecule, have been extensively investigated \cite{R7,R8}.
However, the reverse phenomenon, one single photon simultaneously excited
two atoms or two atoms jointly emitted one single photon, has been rarely
studied. Recently, it is rather remarkable that Luigi Garziano \textit{et al.%
}$\,$\cite{R9} have theoretically demonstrated that one photon can
simultaneously excite two or more atoms in ultrastrong-coupling (USC) regime
\cite{R10,R11,R12,R13} of cavity quantum electrodynamics (QED) \cite{R2}. In
USC regime where the strength of the coupling between the atom and cavity is
comparable to the atom and cavity energy scales, the usual rotating-wave
approximation (RWA) \cite{R14,R15} is no longer valid, and the effect of the
counter-rotating terms becomes important. In this condition, one photon
simultaneously excite two or more atoms via intermediate virtual states
connected by these counter-rotating process, can happen deterministically
\cite{R9}. However, although a few experiments have recently achieved USC
regime in solid-state quantum system \cite{R11,R12,R13}, it is difficult to
manipulate and readout the state of the atom and cavity individually in this
regime with the existing technology, hindering the observation of this novel
phenomenon. In order to get rid of the USC obstacle and excite two atoms
with single photon in the conventional strong coupling regime \cite{R16,R17}, we
investigate a model which combines cavity-assisted Raman process with cavity
photon-mediated interaction.

Commonly, cavity-assisted Raman process employs two field modes interacting
with a three-level atom to induce a two-photon coupling between the field
modes and the atom. The two field modes are dubbed the pump mode and stokes
mode, respectively. This process has been widely studied both theoretically
and experimentally for $\Lambda $-type systems \cite{R18,R19,R20,R21} and $%
\Xi $-type systems \cite{R22,R23}. In general, a Raman interaction
Hamiltonian can be obtained by adiabatically eliminating an auxiliary level
of a three-level system, yielding an effective two-level system with
two-photon coupling \cite{R18,R19,R24}.

On the other hand, it is well known that the exchange of real or virtual
photons between two distant atoms results a photon-mediated interaction,
which can be used as a general tool to distribute quantum information among
different atoms in quantum information processing \cite{R5,R6,R25,R26}.
Recently, with the fast progress of solid-state quantum information
processing \cite{R27,R28,R29}, people have demonstrated the cavity
photon-mediated interaction between two distant artificial atoms \cite%
{R30,R31} in the circuit QED system, which is a solid-state version of
cavity QED. The cavity photon-mediated interaction between artificial atoms
is frequently used to couple two qubits and generate entanglement \cite%
{R32,R33,R34}.

In this work, by combining the cavity-assisted Raman process with cavity
photon-mediated interaction, we propose a new approach to simultaneously
excite two atoms in the strong coupling regime of cavity QED. We consider a
system of a two-level atom and a $\Lambda $-type or V-type three-level atom
off-resonantly coupled to a cavity mode, as shown in Fig. 1. By using a
unitary transformation, we have obtained an effective Hamiltonian that
reflects the essential physics of the process, a single photon can
simultaneously excite two atoms. Furthermore, we generalize above process to
different cases and find that multiple photons and even a classical field can
simultaneously excite two atoms. In addition, we propose a scheme to realize
our proposal in a circuit QED architecture, which consists of two
tunable-gap flux qubits \cite{R35,R36,R37} capacitively coupled to a
superconducting coplanar waveguide resonator \cite{R38}. Using
experimentally feasible parameters, we numerically simulate the quantum
dynamics of this system. It is found that the quantum statistics of this
interesting phenomena may provide insight into the essence of this physical
process. In particular, we discuss the time evolution of equal-time
second-order correlation function in the present work \cite{R9,R39,R40}.

Contrary to that of Luigi Garziano \textit{et al.} \cite{R9} which requires
the ultrastrong coupling, our scheme can be realized in the conventional
strong coupling regime, in which the system can be well described by the
Jaynes-Cummings (JC) Hamiltonian \cite{R14,R15}. Therefore, one can
accurately manipulate the state of the cavity and the atom individually, and
extract the quantum information from this system with high fidelity \cite%
{R29}. By using our scheme, one may easily observe this interesting
phenomena with the state-of-art technology.

The rest of this paper is organized as follows. In Sec.II, we introduce our
model and derive the effective Hamiltonian. In Sec.III, we generalize the
effective Hamiltonian derived in Sec.II, and consider several special cases.
In Sec.IV, we show a scheme to realize our proposal in a circuit QED
architecture, and give the numerical analysis of the dynamics and quantum
statistical properties of the process using experimentally feasible
parameters. In Sec.V, we give conclusions of our investigation and point out
some potential applications.
\begin{figure}[tbp]
\begin{center}
\includegraphics[width=8.0cm,height=4.0cm]{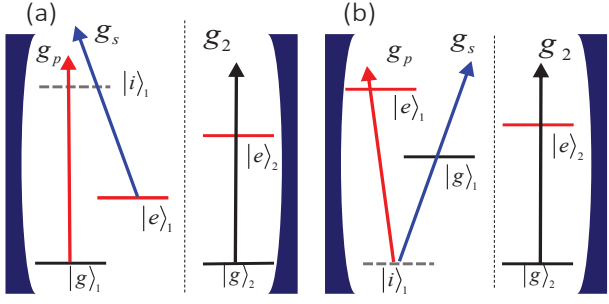}
\end{center}
\caption{(Color online) Energy level diagram of the system. A system of a
two-level atom and a $\Lambda $-type (a) or V-type (b) three-level atom
both off-resonantly coupled to a cavity mode. The arrows denotes the coupling
between the atom and the cavity mode and the gray dashed line represents the
ancillary level in the three-level atom.}
\end{figure}

\begin{figure}[tbp]
\begin{center}
\includegraphics[width=8cm,height=6cm]{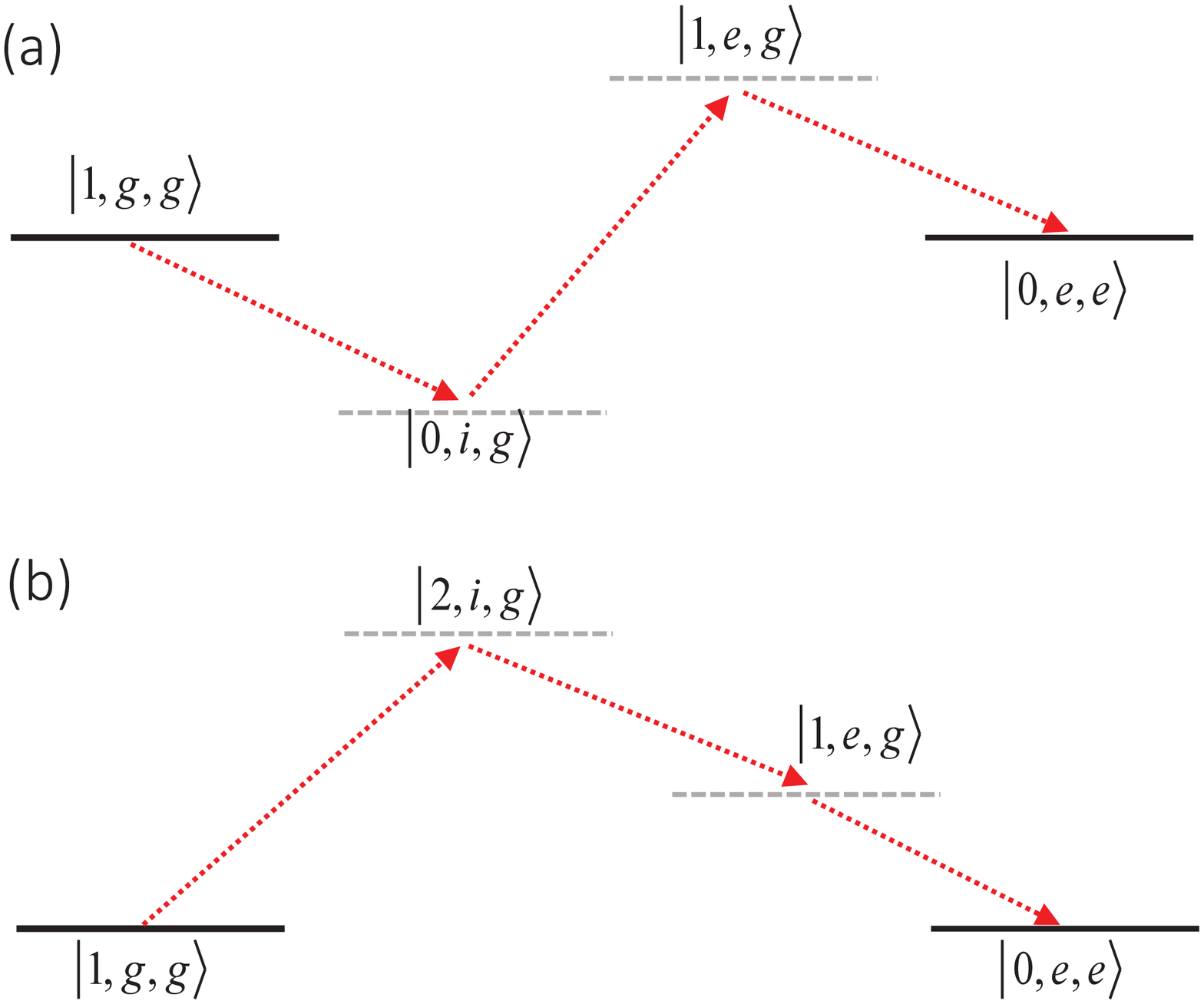}
\end{center}
\caption{(Color online) Sketch of the process in third-order perturbation
theory dominantly contributes to the effective coupling between the bare
states $|1,g,g\rangle $ and $|0,e,e\rangle $. (a) and (b) correspond the
system depicted in Fig. 1(a) and (b), respectively. The red dashed arrows
denote the virtual transitions that doe not conserve the energy, and the gray
dashed lines represent the intermediate virtual states.}
\end{figure}

\section{the model and hamiltonian}

We firstly consider a physical system constituted by a two-level atom and a $%
\Lambda $-type three-level atom as shown in Fig.$\,$1(a). Both atoms are
strongly coupled to a cavity mode. Using the usual RWA, we obtain the
Hamiltonian of this system ($\hbar =1$)
\begin{eqnarray}
\begin{aligned}
H=H_{0}+H_{I},
\end{aligned}
\end{eqnarray}
where
\begin{eqnarray}
\begin{aligned}
H_{0}=\omega_{c}a^{\dagger}a+\sum_{j=g,e,i}\omega_{j}|j\rangle_{1}\langle j|+\omega_{2}\sigma_{2}^{+}\sigma_{2}^{-},
\end{aligned}
\end{eqnarray}
contains the energy of a cavity mode, a $\Lambda $-type three-level atom,
and a two-level atom, respectively. The second term describes the strong
coupling between the two atoms and the cavity mode,
\begin{eqnarray}
\begin{aligned}
H_{I}=g_{p}a|i\rangle_{1}\langle g|+g_{s}a|i\rangle_{1}\langle e|+g_{2}a\sigma_{2}^{+}+H.c.,
\end{aligned}
\end{eqnarray}
here $H.c.$ stands for Hermitian conjugate, $a^{\dagger }$ and $a$ are the creation and annihilation operator for a
cavity mode with frequency $\omega _{c}$, respectively. $\sigma _{2}^{\pm }$ are the
ladder operators for the two-level atom with transition frequency $\omega _{2}
$. $\omega _{j}\,(j=g,e,i)$ is the transition frequency of the three-level
atom from ground to excited state $|j\rangle _{1}$. $g_{k}\,(k=p,s,2)$
denotes the coupling strength between the two atoms and the cavity mode. For simplicity, we
define $\omega _{g}=0$ in the following discussion.

We consider that the system operates in the dispersive regime, where the
atom-cavity detuning is much larger than the coupling strength between
them. Then we have $|\Delta _{p}|=|\omega _{i}-\omega _{c}|\gg g_{p},$ $|\Delta
_{s}|=|\omega _{i}-\omega _{e}-\omega _{c}|\gg g_{s},$ and $|\Delta
_{2}|=|\omega _{2}-\omega _{c}|\gg g_{2}$. The frequency of the cavity mode
satisfies $\omega _{c}\approx \omega _{e}+\omega _{2}$. By using the third-order perturbation theory \cite%
{R9} or a unitary transformation \cite{R20,R41} (The detailed derivation is
given in Appendix A), we can write the effective Hamiltonian for our
system:
\begin{eqnarray}
\begin{aligned}
H_{eff}=&\omega_{c}a^{\dagger}a+\omega_{1} \sigma_{1}^{+}\sigma_{1}^{-}+\omega_{2} \sigma_{2}^{+}\sigma_{2}^{-}
\\&+( \chi a \sigma_{1}^{+} \sigma_{2}^{+} + H.c.),
\end{aligned}
\end{eqnarray}
where
\begin{eqnarray}
\begin{aligned}
\chi=\frac{g_{s}g_{2}}{3}(\frac{1}{\Delta_{s}} + \frac{1}{\Delta_{2}})\frac{g_{p}}{\Delta_{p}} + \frac{g_{p}g_{s}}{3}(\frac{1}{\Delta_{p}} + \frac{1}{\Delta_{s}})\frac{g_{2}}{\Delta_{2}},
\end{aligned}
\end{eqnarray}
$\sigma _{1}^{\pm }$ are the ladder operators for a
reduced two-level system of frequency $\omega _{1}\,(\omega _{e})$ formed by
the lowest two levels ($|g\rangle _{1}$ and $|e\rangle _{1}$) of the $\Lambda $%
-type three-level atom. For simplicity, we have omit the term which
constitutes a renormalization of the two atoms and cavity energy levels in
Eq.$\,$(4) and also throughout the rest of the main text (see the Appendix
A, this renormalization of the energy level is a result of dispersive
coupling between the atom and the cavity mode). In addition, since the
system is operated in the dispersive regime, it is a good approximation to still use
the the bare states \cite{R59}, which are the eigenstates of the uncoupled Hamiltonian $%
H_{0}$, instead of the transformed basis (dressed-state basis). We will
use this approximation in whole paper. For a physical system constituted by a two-level atom and a V-type
three-level atom as shown in Fig.$\,$1(b), we can use same procedure and
obtain similar results (see Appendix B).

The last term in Eq.$\,$(4), $\chi a\sigma _{1}^{+}\sigma _{2}^{+}+H.c.$,
describes a coherent coupling between one photon and two atoms with strength
$\chi $. This implies that one can simultaneously excite two atoms with one
single cavity photon. As shown in Fig.$\,$2(a), we present the whole process
in third-order perturbation theory which leads to the effective coupling
between the bare states $|1,g,g\rangle $ and $|0,e,e\rangle .$ For the state
vector of the bare state $\left\vert n,j,k\right\rangle $, the first number
denotes the photon number state of the cavity, the second and third entries
denote the state of the three-level atom and the two-level atom,
respectively. The transition between $|1,g,g\rangle $ and $|0,e,e\rangle $
is connected by three virtual transitions. Fig.$\,$2(b)
shows the process for the system depicted in Fig.$\,$1(b).

Although one can obtain the effective coupling between  $|1,g,g\rangle $ and
$|0,e,e\rangle $ from the perturbation theory, two alternative explanations
based on Raman-type process and cavity photon-mediated interaction can better
illustrate the mechanism of the process.

(1) The $\Lambda $-type three-level atom is off-resonantly coupled to the
cavity mode. This leads to an effective two-photon Raman coupling $%
a^{\dagger }a(\sigma _{1}^{+}+\sigma _{1}^{-})$ \cite{R42}. The coupling
strength $J_{a}\sim g_{p}g_{s}(\frac{1}{\Delta _{p}}+\frac{1}{\Delta _{s}})$%
. Furthermore, exchanging a virtual photon between the reduced two-level
atom and the second two-level atom results an effective Hamiltonian $%
(a\sigma _{1}^{+}\sigma _{2}^{+}+H.c.)$ with coupling strength $\chi
_{a}\sim \frac{J_{a}g_{2}}{\Delta _{2}}$.

(2) The cavity photon-mediated interaction between the two atoms results an
effective coupling ($a^{\dagger }|g\rangle _{1}\langle i|+H.c.$) \cite%
{R30,R41,R43} with coupling strength $J_{b}\sim g_{s}g_{2}(\frac{1}{\Delta
_{s}}+\frac{1}{\Delta _{2}})$. For the $\Lambda $-type three-level atom, we
may think this cavity photon-mediated coupling as atomic-type Stokes mode,
and the cavity mode as the pump mode. Therefore, when the three-level system
is driven by the two modes, we can adiabatically eliminate the auxiliary
level of the three-level atom. Now the three-level atom reduces to an
effective two-level atom, and an effective coupling between the two atoms
and the cavity mode $(a\sigma _{1}^{+}\sigma _{2}^{+}+H.c.)$ is obtained.
The coupling strength is $\chi _{b}\sim \frac{J_{b}g_{p}}{\Delta _{p}}$.

The above two different approaches both include a Raman-type process assisted by
the ancillary level $|i\rangle _{1}$ and cavity photon-mediated coupling. It
is apparently that the two effective coupling strength $\chi _{a}$ and $\chi
_{b}$ correspond to the two terms in Eq.$\,$(5). From this qualitative
viewpoint, the above process can be easily generalized to the various cases, as shown in the following section.

\section{simultaneously exciting two atoms: general formulation}

In this section, we give a general formulation for the process of
simultaneously exciting two atoms.

The physics of a three-level atom interacting with two field modes (quantum
or classical) were studied extensively \cite{R15,R44}. In these systems,
cavity-assisted Raman process can leads to two-photon coupling. The
Hamiltonian is given by $(a_{p}^{\dagger }a_{s}\sigma ^{-}+H.c.)$ for $%
\Lambda $- or V-type atoms \cite{R21} and $(a_{p}^{\dagger }a_{s}^{\dagger
}\sigma ^{-}+H.c.)$ for $\Xi $-type atoms \cite{R22}, where $\sigma ^{\pm }$
are the ladder operator for the reduced two-level system, $a_{p}$ and $a_{s}$
are the annihilation operators for the two field modes (the pump mode and the
stokes mode), respectively. For the case of classical field mode, one can simply replace
the $a_{p}$ or $a_{s}$ with $\epsilon e^{-i(\omega _{d}t+\phi )}$ , where $%
\epsilon $ and $\phi $ are the real amplitude and phase of the classical
field at frequency $\omega _{d}$.

Combining these Raman-assisted two-photon coupling with cavity
photon-mediated interaction, we can obtain a general approach to simultaneously
excite two atoms.

\subsection{two photons simultaneously excite two atoms}

Here, we consider the process of two photons exciting two atoms. By jointly
absorbing two photons, two atoms are excited from their ground state to the
excited state. We noted that in the whole process the single photon can not
be absorbed by any one of the two atoms. This resonant interaction between
two photons and two atoms can be described with the Hamiltonian (in the
interaction picture, and $\hbar =1$),
\begin{eqnarray}
\begin{aligned}
H^{I}= \chi_{2} (a)^{2} \sigma_{1}^{+} \sigma_{2}^{+} + H.c.,
\end{aligned}
\end{eqnarray}
where $a$\thinspace $(a^{\dagger })$ is the photon annihilation (creation)
operator, $\sigma _{q}^{\pm }(q=1,2)$ are the ladder operators for the $q$th atom, and $\chi _{2}$ denotes the coupling strength.

This effective model can be realized in a system consisting of two three-level atoms.
One is $\Lambda $-type and we denote the states as $|g\rangle _{1},|e\rangle
_{1},$and $|i\rangle _{1}$. The other is $\Xi $-type and we denote the
states as $|g\rangle _{2},|i\rangle _{2},$and $|e\rangle _{2}$. Both atoms
are dispersively coupled to a cavity mode. Under the frequency matching
condition, by using the fourth-order perturbation theory \cite{R9}, we can
obtain an effective interaction Hamiltonian, which can be described by Eq.$\,
$(6). According to fourth-order perturbation theory, the transition between $|2,g,g\rangle $ and $|0,e,e\rangle $ is
enabled by paths with four virtual transitions. One trivial example of such path is $|2,g,g\rangle \longrightarrow
|1,i,g\rangle \longrightarrow |2,e,g\rangle \longrightarrow |2,e,i\rangle
\longrightarrow |2,e,e\rangle $.

Following the same procedure as that used in Sec.$\,$II, we give a
qualitative explanation of the above effective coupling instead of a full
analytical derivation. Firstly, for the two three-level atoms, cavity-assisted
Raman process can leads to two-photon coupling interaction, which is given by
$a^{\dagger }a(\sigma _{1}^{-}+\sigma _{1}^{+})$ for the $\Lambda $-type
atom and $(a^{\dagger })^{2}\sigma _{2}^{-}+(a)^{2}\sigma _{2}^{+}$ for the $%
\Xi $-type atom.  Here $\sigma _{1}^{\pm }$ are the ladder operators for a
reduced two-level system formed by the lowest two levels $|g\rangle _{1}$
and $|e\rangle _{1}$ of the $\Lambda $-type three-level atom. $\sigma
_{2}^{\pm }$ are the ladder operators acting on the ground and second excited
states ($|g\rangle _{2}$ and $|e\rangle _{2}$) of the $\Xi $-type three-level
atom. It is apparently that the two reduced two-level atoms are both coupled to a
common cavity mode. By exchanging virtual photon with the cavity, a effective coupling between two atoms
and two photons is obtained, which can be described by $H_{I}$
in Eq.$\,$(6). In fact, since the number of the excitations is conserved in
this case, this effective coupling can also be realized in two-atom
Tavis-Cummings system\cite{R46} operating in the dispersive regime.
According to the second-order perturbation theory, the effective coupling
between the states $|2,g,g\rangle $ and $|0,e,e\rangle $ is enabled by paths
with two virtual transitions. For example, one of such a path is $%
|2,g,g\rangle \longrightarrow |1,g,e\rangle \longrightarrow |0,e,e\rangle $.

For multi-photon transitions, we can simply replace $(a^{\dagger
})^{2}\sigma _{2}^{-}+(a)^{2}\sigma _{2}^{+}$ with $(a^{\dagger })^{n}\sigma
_{2}^{-}+(a)^{n}\sigma _{2}^{+}$. Therefore, we can easily extend the
process to the multi-photon case and obtain the effective coupling $%
a^{n}\sigma _{1}^{+}\sigma _{2}^{+}+H.c$.

\subsection{a class field simultaneously excite two atoms}

From Eq.$\,$(4), we know that in the interaction picture the general
Hamiltonian for simultaneously exciting two atoms with one photon is
\begin{eqnarray}
\begin{aligned}
H^{I}= \chi (a \sigma_{1}^{+} \sigma_{2}^{+} + H.c.).
\end{aligned}
\end{eqnarray}
In the parametric approximation, the cavity mode is treated as a classical
field without depletion. Thus, by replacing $a$ with $\epsilon e^{-i\phi }$,
we transform the Hamiltonian to
\begin{eqnarray}
\begin{aligned}
H^{I}= \chi'(\sigma_{1}^{+} \sigma_{2}^{+}e^{-i\phi} + H.c.),
\end{aligned}
\end{eqnarray}
where the effective coupling strength is $\chi ^{\prime }=\chi \epsilon $, $%
\epsilon $ and $\phi $ are the real amplitude and phase of the classical
field. For simplicity, we set $\phi =0$ in the following discussion.

This effective model can be realized in a physical system depicted in Fig.$\,
$1. However, we need to apply a classical driving field on the three-level
atom. This classical driving field is introduced as a classical pump mode. For
example, we consider a system depicted in Fig.$\,$1(a). When we apply a
classical driving field only on the $\Lambda $-type three-level atom, the
Hamiltonian of the system is
\begin{eqnarray}
\begin{aligned}
&H=H_{0}+H_{int},
\\&H_{int}=H_{I}+H_{D}
\end{aligned}
\end{eqnarray}
with $H_{0}$\,$(H_{I})$ being given in Eq.$\,$(2)$\,$(Eq.$\,$(3)). The third term
comes from the drive of the $\Lambda $-type three-level atom,
\begin{eqnarray}
\begin{aligned}
H_{D}=\varepsilon e^{-i\omega_{d}t}|i\rangle_{1}\langle g|+\varepsilon' e^{-i\omega_{d}t}|i\rangle_{1}\langle e|+H.c.,
\end{aligned}
\end{eqnarray}
where $\varepsilon $ and $\varepsilon ^{\prime }$ are the real amplitude of
the classical driving field with frequency $\omega _{d}$. Then we can rearrange the terms in $H_{int}$, and rewrite $H_{int}$ as
\begin{eqnarray}
\begin{aligned}
H_{int}&=(\varepsilon e^{-i\omega_{d}t}|i\rangle_{1}\langle g|+g_{s}a|i\rangle_{1}\langle e|+g_{2}a\sigma_{2}^{+}+H.c.)
\\&+(g_{p}a|i\rangle_{1}\langle g|+\varepsilon' e^{-i\omega_{d}t}|i\rangle_{1}\langle e|+H.c.),
\end{aligned}
\end{eqnarray}
where the terms in the first line have the same form as that of the Hamiltonian in Eq.$\,$(3) except replacing the quantum pump mode $(g_{p}a|i\rangle_{1}\langle g|+H.c.)$ with a classical pump mode $(\varepsilon e^{-i\omega_{d}t}|i\rangle_{1}\langle g|+H.c.)$. The terms in the second line have no contribution to the process except shifting the frequency of the energy level in our setting. By analogy with the mechanism discussed in last section, we can simultaneously excite two atoms with a classical driving field.

We still consider that the atom-cavity system operates in the dispersive
regime, and the $\Lambda $-type three-level system is off-resonantly driven
by the classical field. Then $|\Delta _{d}|=|\omega _{i}-\omega _{d}|\gg
\varepsilon $ and $|\Delta _{d}^{\prime }|=|\omega _{i}-\omega _{e}-\omega
_{d}|\gg \varepsilon ^{\prime }$. The frequency of the classical driving
field satisfies $\omega _{d}\approx \omega _{e}+\omega _{2}$. Following the
derivation in Sec.$\,$II, we can write the effective Hamiltonian in the
interaction picture,
\begin{eqnarray}
\begin{aligned}
H^{I}_{d}= \chi_{d}(\sigma_{1}^{+} \sigma_{2}^{+} + H.c.),
\end{aligned}
\end{eqnarray}
where $\sigma _{1}^{\pm }$ are the ladder operators for a reduced two-level
system with frequency $\omega _{1}$ formed by the lowest two levels of the $%
\Lambda $-type three-level atom. The effective coupling strength is
\begin{eqnarray}
\begin{aligned}
\chi_{d}=\frac{g_{s}g_{2}}{3}(\frac{1}{\Delta_{s}} + \frac{1}{\Delta_{2}})\frac{\varepsilon}{\Delta_{d}} + \frac{\varepsilon g_{s}}{3}(\frac{1}{\Delta_{d}} + \frac{1}{\Delta_{s}})\frac{g_{2}}{\Delta_{2}}.
\end{aligned}
\end{eqnarray}

This classical field-induced amplitude- and phase-tunable two-atom coupling
can be dubbed the two-photon coherent pump \cite{R45}. When the system is
initially prepared in the state $|g,g\rangle$, the two-atom GHZ state $%
(|g,g\rangle+|e,e\rangle)/\sqrt2$ can be obtained by applying the classical
field on the three level atom after a time $t=\pi/(4\chi_{d})$.

\section{physical implementation}

With the recent rapid progress in quantum information processing,
investigation has been extended from cavity QED systems to circuit QED
systems, in which superconducting qubits \cite{R47} act as artificial atoms
and can strongly interact with a single-mode field at microwave frequencies.
The artificial atoms with a long coherence time can be engineered to have
different energy level diagrams, including V-type \cite{R48,R49}, $\Delta $%
-type \cite{R37}, and $\Xi $-type \cite{R50,R51}. The transition frequency of the atoms
can be controlled by a local magnetic flux bias. On the other hand, the
superconducting resonator with high quality factor can be easily designed
and fabricated with the existing technology. Moreover, one can manipulate
the state of the resonator and the artificial atom individually, and read
out the states of these system with high fidelity \cite{R29}. These features
along with the potential of scaling up make the circuit-QED system an
attractive platform for studying quantum optics and quantum information
processing.

In this section, we show that a single cavity photon can simultaneously
excite two atoms in a circuit QED architecture. We numerically simulate the
dynamics and quantum statistical properties of the process using
experimentally feasible parameters. The numerical calculation were performed
using the PYTHON package QuTiP \cite{R52,R53}.

\begin{figure}[tbp]
\begin{center}
\includegraphics[width=8.0cm,height=6.0cm]{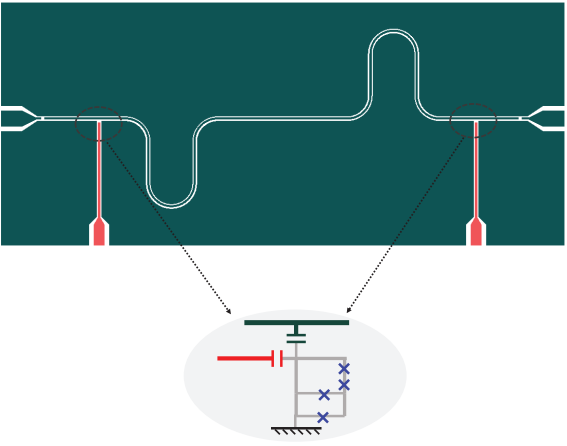}
\end{center}
\caption{(Color online) Schematic of the superconducting system consisting
of two tunable-gap flux qubits with local microwave driving lines (light gray, red online),
which are both located in a transmission line resonator (dark gray, green online).}
\end{figure}

\subsection{implementation in circuit quantum electrodynamics}

\begin{figure}[tbp]
\begin{center}
\includegraphics[width=8.0cm,height=4.0cm]{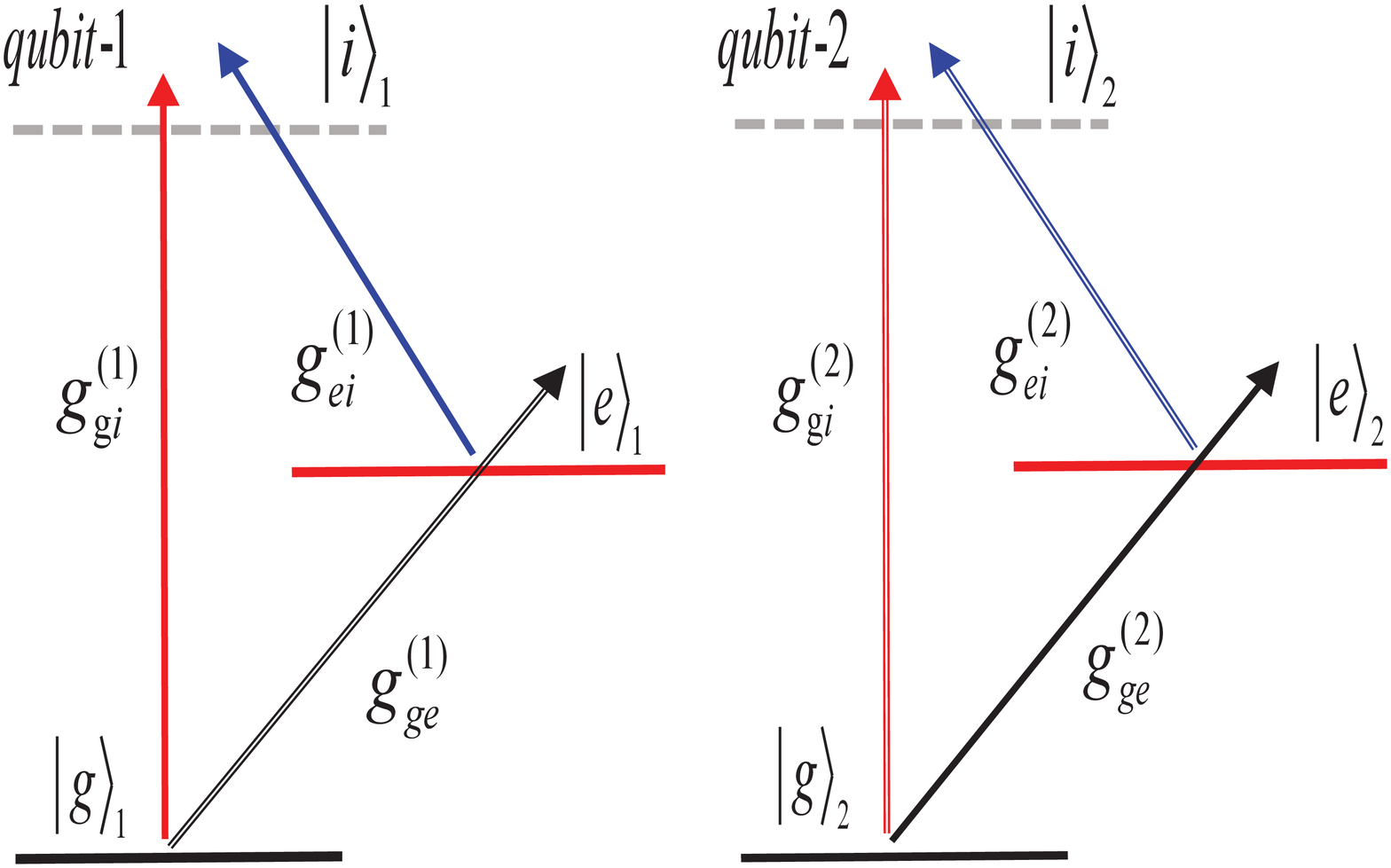}
\end{center}
\caption{(Color online) Energy level diagram of the two flux qubits. The $%
\Delta $-type three-level system is off-resonantly coupled to the cavity
mode. The present system, which is formed by a pair of identical units
(represented by single line and double line, respectively), can be seen as a
dimer of the system described in Fig.$\,$1(a). The gray dashed line
represents the ancillary level in the three-level system.}
\end{figure}

As shown in Fig.$\,$3, we consider two tunable-gap flux qubits capacitively
coupled to a superconducting coplanar waveguide resonator \cite{R36,R37} in
the strong coupling regime. The flux qubit can be manipulated by a local
microwave driving line. To initialized the state of the cavity, we can add
an additional frequency tunable qubit (not shown in Fig.$\,$3), which allows
us to pump photon into the resonator \cite{R54,R55,R56}. Treating the
tunable-gap flux qubit as a $\Delta $-type three-level system and using the
usual RWA, we can describe the full system with Hamiltonian ($\hbar =1$)
\begin{eqnarray}
\begin{aligned}
&H=\omega_{c}a^{\dagger}a+\sum_{q=1,2}\sum_{j=g,e,i}\omega^{(q)}_{j}|j\rangle_{q}\langle j|+H_{I},
\\&H_{I}=\sum_{q=1,2}a(g^{(q)}_{ge}|e\rangle_{q}\langle g|+g^{(q)}_{gi}|i\rangle_{q}\langle g|+g^{(q)}_{ei}|i\rangle_{q}\langle e|)+H.c,
\end{aligned}
\end{eqnarray}
where $\omega _{c}$ is cavity frequency, $\omega _{j}^{(q)}$ is the
transition frequency from ground state to excited state $|j\rangle _{q}$ of
the $q$th flux qubit, and $g_{jk}^{(q)}$ is the qubit-resonator coupling
strength for the $q$th flux qubit. For easy reference, we define $\omega
_{g}^{(q)}=0$ in the following discussion. We assume that our system
operates in the dispersive regime. Then $|\Delta _{2}^{(q)}|=|\omega
_{e}^{(q)}-\omega _{c}|\gg g_{ge}^{(q)}$, $|\Delta _{s}^{(q)}|=|\omega
_{i}^{(q)}-\omega _{e}^{(q)}-\omega _{c}|\gg g_{ei}^{(q)}$, and $|\Delta
_{p}^{(q)}|=|\omega _{i}^{(q)}-\omega _{c}|\gg g_{gi}^{(q)}$. The frequency
of the cavity satisfies the frequency matching condition $\omega _{c}\approx
\sum_{q=1,2}\omega _{e}^{(q)}$.

As shown in Fig.$\,$4, the present system can be seen as a dimer of the
system depicted in Fig.$\,$1(a). Writing out all terms of the summation and
rearranging them, one can rewrite $H_{I}$ to
\begin{eqnarray}
\begin{aligned}
H_{I}&=(g^{(1)}_{gi}a|i\rangle_{1}\langle g|+g^{(1)}_{ei}a|i\rangle_{1}\langle e|+g^{(2)}_{ge}a|e\rangle_{2}\langle g|+H.c.)
\\&+(g^{(2)}_{gi}a|i\rangle_{2}\langle g|+g^{(2)}_{ei}a|i\rangle_{2}\langle e|+g^{(1)}_{ge}a|e\rangle_{1}\langle g|+H.c.)
\end{aligned}
\end{eqnarray}
The terms in the first and second line both have the same form as Eq.$\,$(3). By analogy with the system discussed in Sec.$\,$II, one can also simultaneously excite two atoms with a single cavity
photon in this Circuit-QED system. Moreover, it is apparently that except
for a path which is the same as that shown in Fig.$\,$2(a), there is another
path generating the effective coupling between the states $|1,g,g\rangle $
and $|0,e,e\rangle $ after considering the third-order perturbation. The
path is $|1,g,g\rangle \longrightarrow |0,g,i\rangle \longrightarrow
|1,g,e\rangle \longrightarrow |0,e,e\rangle $, where $\left\vert
n,j,k\right\rangle $ labels the states of the cavity mode and two three-level systems.

To take into account that our system operates in the dispersive regime, we can eliminate the direct atom-cavity coupling
by using the unitary transformation
\begin{eqnarray}
\begin{aligned}
U=\exp[\sum_{q=1,2}&\frac{g_{gi}^{(q)}}{\Delta_{p}^{(q)}}a^{\dagger}\left|g\right\rangle_{q}\left\langle i\right|+\frac{g_{ei}^{(q)}}{\Delta_{s}^{(q)}}a^{\dagger}\left|e\right\rangle_{q}\left\langle i\right|\\&+\frac{g_{ge}^{(q)}}{\Delta_{2}^{(q)}}a^{\dagger}\left|g\right\rangle_{q}\left\langle e\right|-H.c.]
\end{aligned}
\end{eqnarray}
Following the derivation in Sec.$\,$II, we can write the effective
Hamiltonian,
\begin{eqnarray}
\begin{aligned}
H_{eff}=&\omega_{c}a^{\dagger}a+\omega_{1} \sigma_{1}^{+}\sigma_{1}^{-}+\omega_{2} \sigma_{2}^{+}\sigma_{2}^{-}
\\&+( \chi_{eff} a \sigma_{1}^{+} \sigma_{2}^{+} + H.c.),
\end{aligned}
\end{eqnarray}
where
\begin{eqnarray}
\begin{aligned}
\chi_{eff}=&\frac{g^{(1)}_{ei}g^{(2)}_{ge}}{3}(\frac{1}{\Delta_{s}^{(1)}} + \frac{1}{\Delta_{2}^{(2)}})\frac{g^{(1)}_{gi}}{\Delta_{p}^{(1)}}
\\&+ \frac{g^{(1)}_{gi}g^{(1)}_{ei}}{3}(\frac{1}{\Delta_{p}^{(1)}} + \frac{1}{\Delta_{s}^{(1)}})\frac{g^{(2)}_{ge}}{\Delta_{2}^{(2)}}
\\&+\frac{g^{(2)}_{ei}g^{(1)}_{ge}}{3}(\frac{1}{\Delta_{s}^{(2)}} + \frac{1}{\Delta_{2}^{(1)}})\frac{g^{(2)}_{gi}}{\Delta_{p}^{(2)}}
\\&+ \frac{g^{(2)}_{gi}g^{(2)}_{ei}}{3}(\frac{1}{\Delta_{p}^{(2)}} + \frac{1}{\Delta_{s}^{(2)}})\frac{g^{(1)}_{ge}}{\Delta_{2}^{(1)}} ,
\end{aligned}
\end{eqnarray}
$\sigma _{q}^{\pm }$ are the ladder operators for a qubit formed by the lowest two levels $|g\rangle_{q} $ and $|e\rangle _{q}$ of the $q$th three-level system. The frequency of the qubit is $\omega
_{q}.$

\subsection{numerical analysis}

\begin{figure}[tbp]
\begin{center}
\includegraphics[width=8.0cm,height=6.0cm]{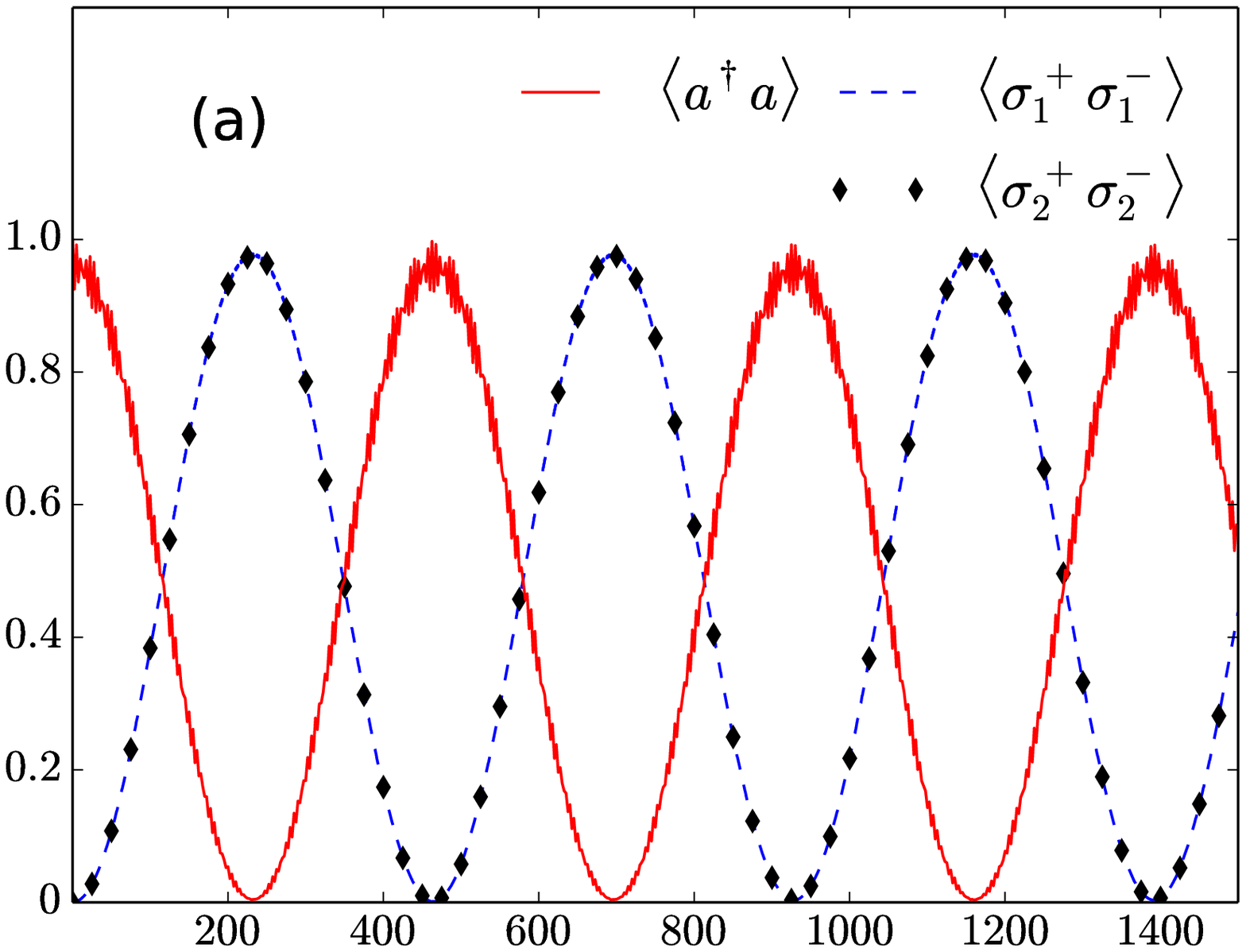}
\includegraphics[width=8.0cm,height=2.0cm]{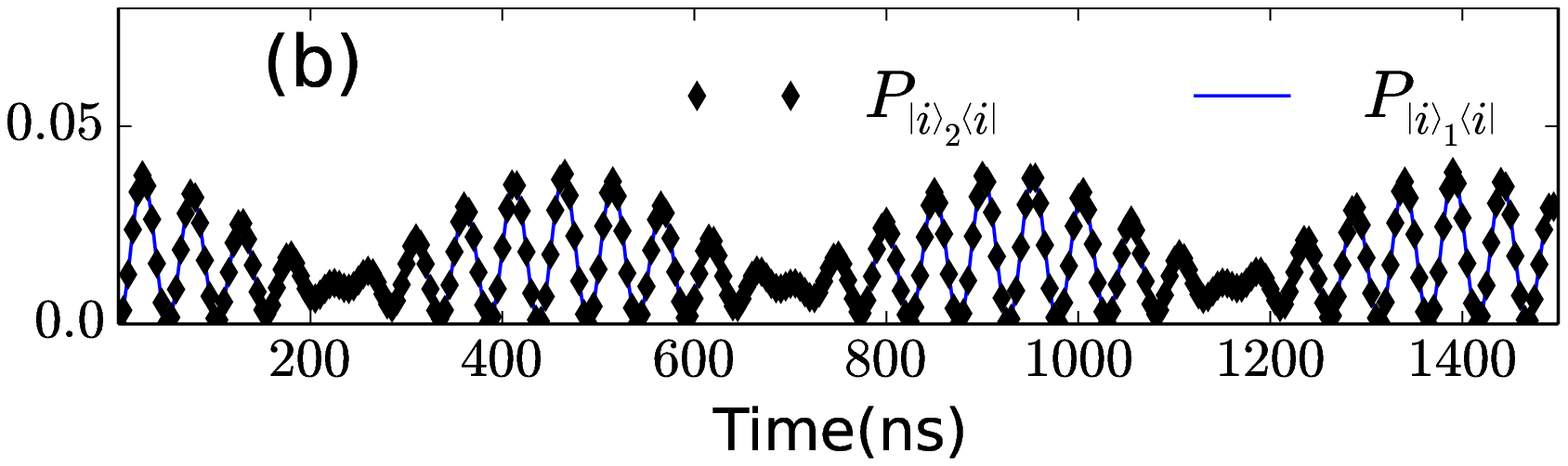}
\end{center}
\caption{(Color online) Numerical simulation of the nondissipative dynamics.
(a) Temporal evolution of the cavity mean photon number $\langle a^{\dagger
}a\rangle $ and the qubit mean excitation number $\langle \protect\sigma %
_{q}^{+}\protect\sigma _{q}^{-}\rangle $ for the initial states $\left\vert
1,g,g\right\rangle $. It can be found that the system undergoes ordinary
vacuum Rabi oscillations between $\left\vert 1,g,g\right\rangle $ and $%
\left\vert 0,e,e\right\rangle $, showing the coherent and reversible
excitation exchange between the two qubits and a cavity mode with
probability approaching one. This demonstrates that a single photon is
absorbed and emitted by the two qubits jointly in a reversible and coherent
way. (b) Population leakage to the auxiliary third level of the two three-level systems.
In the whole process, the real occupation of the third level is far less
than one.}
\end{figure}

\begin{figure}[tbp]
\begin{center}
\includegraphics[width=8.0cm,height=6.0cm]{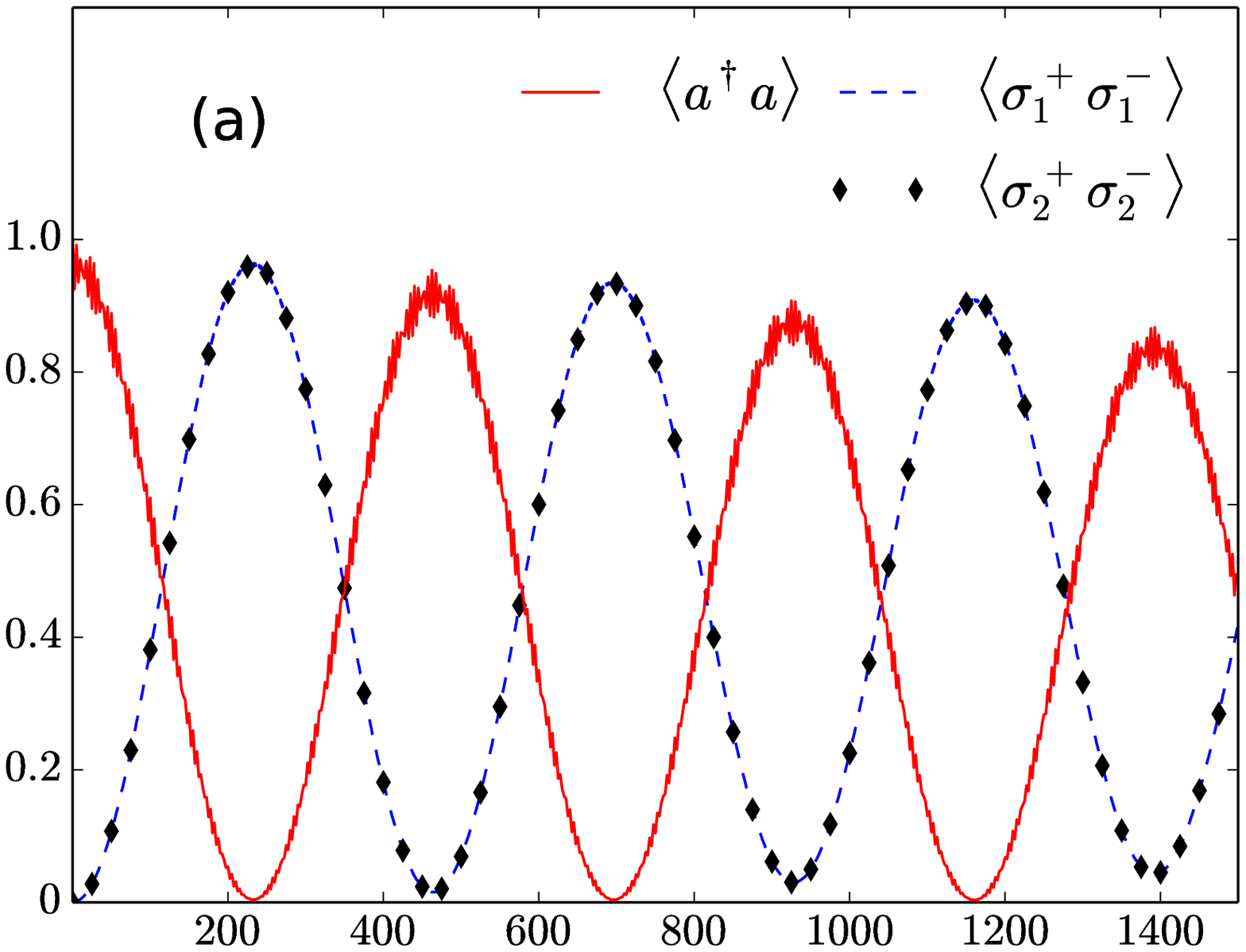} %
\includegraphics[width=8.0cm,height=2.0cm]{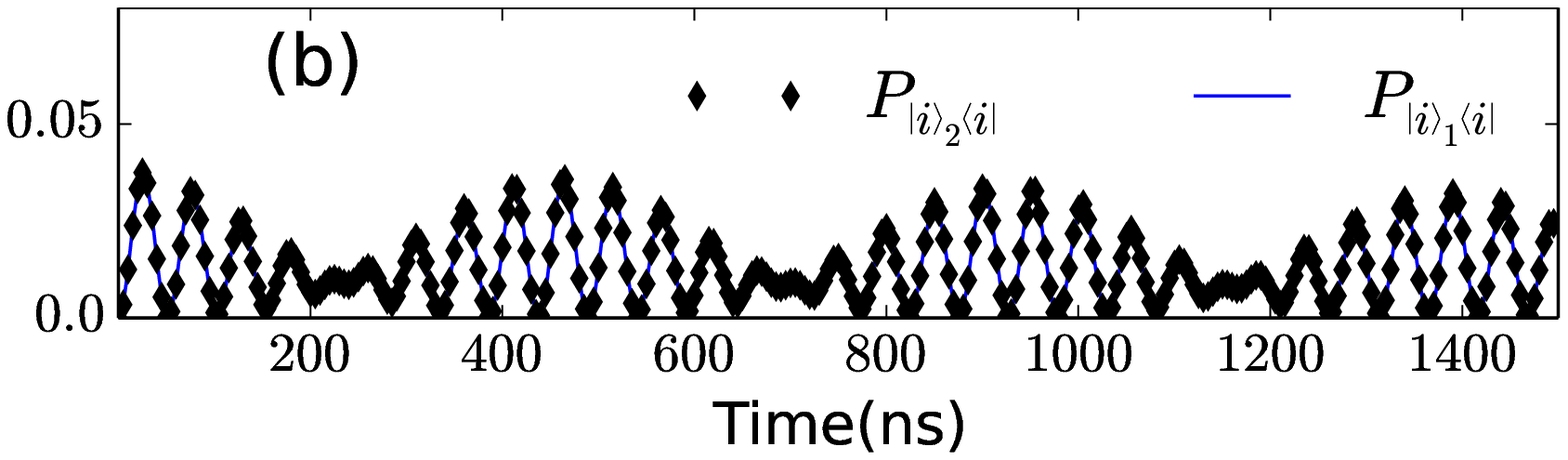}
\end{center}
\caption{(Color online) Numerical simulation of the dynamics under the
influence of dissipation. (a) Temporal evolution of the cavity mean photon
number $\langle a^{\dagger }a\rangle $ and the qubit mean excitation number $%
\langle \protect\sigma _{q}^{+}\protect\sigma _{q}^{-}\rangle $ for the
initial states $\left\vert 1,g,g\right\rangle $. (b) Population leakage to
the auxiliary third level of the two three-level systems.}
\end{figure}

To study the feasibility of the proposal, we have performed numerical
calculations based on the Hamiltonian in Eq.$\,$(14). Note that in the
previous effective Hamiltonian in Eq.$\,$(17), we do not include analytical expressions for the
frequency shifts due to the renormalization and modification caused by the
dispersive coupling between the qubit and the cavity mode. However, we take
them into account by numerically scanning the cavity frequency to compensate
these frequency shift. Therefore, the frequency matching condition can be
fulfilled.

For simplicity and without loss of generality, we consider the two
tunable-gap flux qubits as two identical $\Delta $-type three-level systems.
In what follows, all the numerical analysis are done with the frequencies of
cavity and flux qubit, $\omega _{c}/2\pi =7.9655\,GHz$, $\omega _{e}^{(q)}/2\pi
=4.00\,GHz$, and $\omega _{i}^{(q)}/2\pi =7.00\,GHz$. The coupling strength
are $g_{ge}^{(q)}/2\pi =120\,MHz$, $g_{ei}^{(q)}/2\pi =180\,MHz$, and $%
g_{gi}^{(q)}/2\pi =100\,MHz$. The cavity photon decay rates and the flux qubit
relaxation rates are $\kappa /2\pi =\gamma _{ge}^{(q)}/2\pi =\gamma
_{gi}^{(q)}/2\pi =0.01\,MHz$, and $\gamma _{ei}^{(q)}/2\pi =0.015\,MHz$,
respectively. Further, we also consider that our system is initially
prepared in the state $\left\vert 1,g,g\right\rangle $.

Firstly, we study the dynamics of the system in the absence of dissipation.
In Fig.$\,$5, we present the time evolution of the mean photon number $%
\langle a^{\dagger }a\rangle $ and the qubit mean excitation number $\langle
\sigma _{q}^{+}\sigma _{q}^{-}\rangle $. It can be observed from this
ordinary oscillation between $\left\vert 1,g,g\right\rangle $ and $%
\left\vert 0,e,e\right\rangle $ that a single photon is absorbed and emitted
by the two qubits jointly with probability approaching one. The period of the
oscillation is about 460 $ns$, which is in good agreement with the value
calculated based on the effective coupling strength $\chi _{eff}$, $T=\pi
/\chi _{eff}=444$ $ns$. Fig.$\,$5(b) shows the population leakage of
the auxiliary third level of the two three-level systems. We can find that in the
whole process the real occupation of the third level is far less than one.

We now discuss the influence of cavity decay and qubit relaxation on this
ongoing physical process. This can be done by solving the master equation
(see the Appendix C). In Fig.$\,$6, as expected, the amplitude of the
oscillation gets damped due to cavity decay and qubit relaxation.
However, one can still observe the oscillation between $\left\vert
1,g,g\right\rangle $ and $\left\vert 0,e,e\right\rangle $.

\begin{figure}[tbp]
\begin{center}
\includegraphics[width=8.0cm,height=6.0cm]{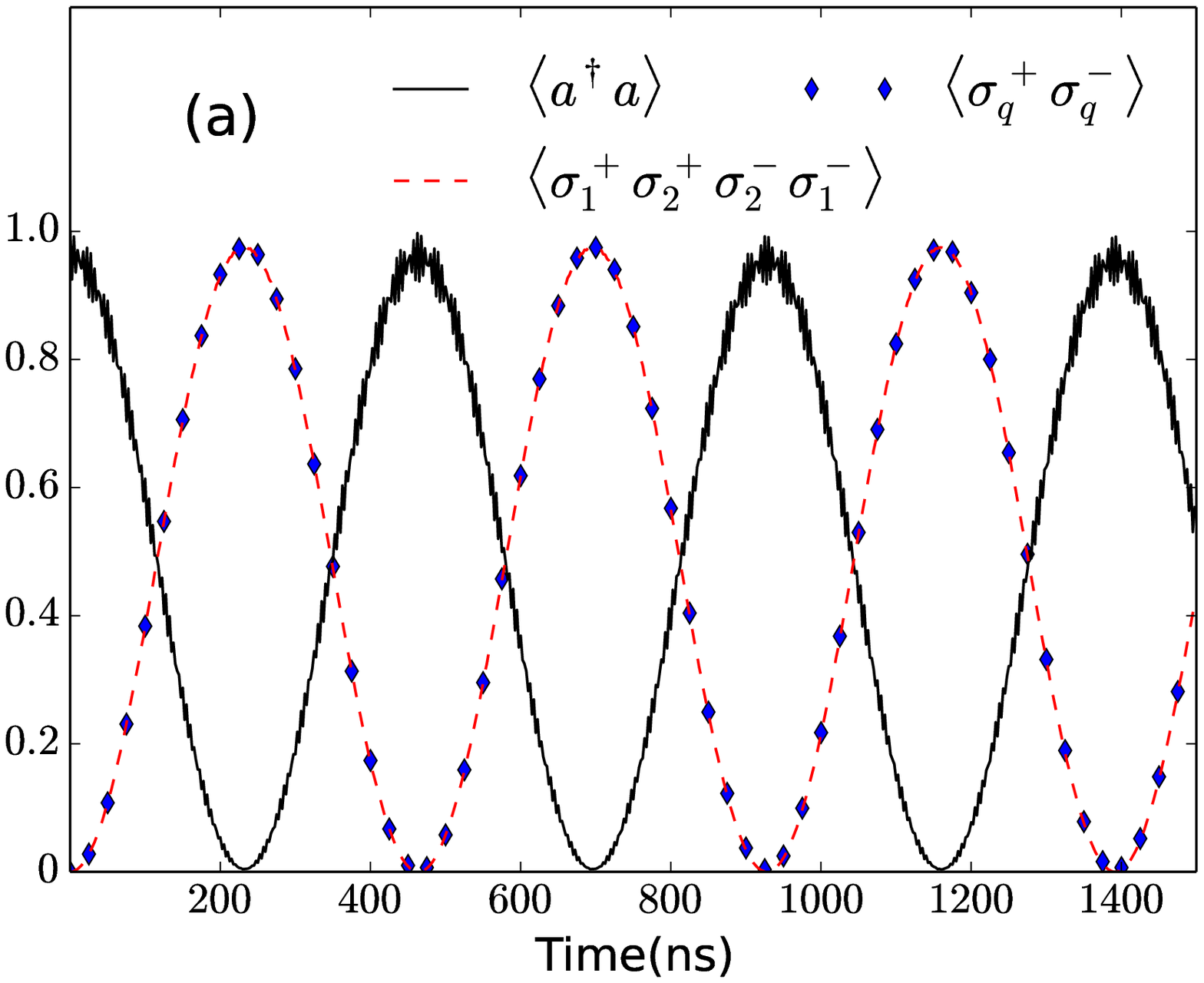} %
\includegraphics[width=8.0cm,height=6.0cm]{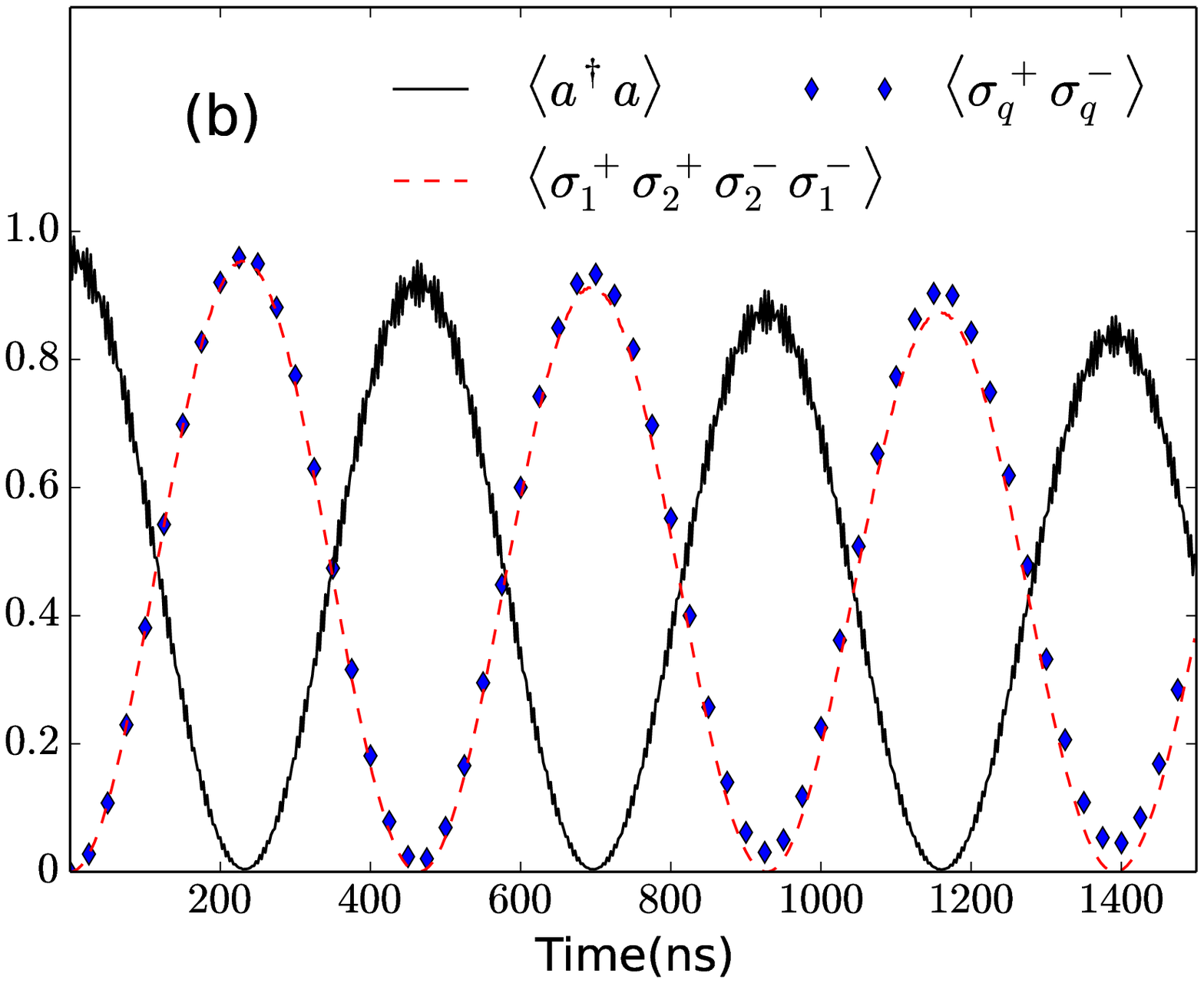}
\end{center}
\caption{(Color online) Temporal evolution of the the qubit mean excitation
number $\langle \protect\sigma _{q}^{+}\protect\sigma _{q}^{-}\rangle $, and
the equal-time second-order correlation function $G_{q}^{(2)}=\langle
\protect\sigma _{1}^{+}\protect\sigma _{2}^{+}\protect\sigma _{2}^{-}\protect%
\sigma _{1}^{-}\rangle $ with the system prepared in the state $\left\vert
1,g,g\right\rangle $. (a) shows dynamics in the absence of dissipation and
(b) for dynamics under the influence of dissipation. In the nondissipative
case, the $\langle \protect\sigma _{q}^{+}\protect\sigma _{q}^{-}\rangle $
and $G_{q}^{(2)}$ almost coincide. This almost perfect two-qubit correlation
function is a direct signature that the two qubits are excited jointly. As
expected, the oscillation amplitude gets damped under the influence of
dissipation. We also observed that the two-qubit correlation function $%
G_{q}^{(2)}$ goes almost to zero when the cavity mean photon number $\langle
a^{\dagger }a\rangle $ is at maximum. This behavior indicates that the
photon is not absorbed by a single qubit instead the two qubits jointly.}
\end{figure}

Most importantly, studying the quantum statistics of this interesting
phenomena may provide insight into the essence of this physical process. We
particularly focus on the time evolution of equal-time second-order
correlation function. In Fig.$\,$7, we display the numerical result of the
time evolution of the two-qubit correlation function $G_{q}^{(2)}=\langle
\sigma _{1}^{+}\sigma _{2}^{+}\sigma _{2}^{-}\sigma _{1}^{-}\rangle $, which
describes the quantum correlation between the emitted photons from the two
qubits into noncavity modes \cite{R9}. For easy comparison, we also display
the numerical result of the time evolution of the qubit mean excitation
number $\langle \sigma _{q}^{\dagger }\sigma _{q}^{-}\rangle $ and the mean
photon number $\langle a^{\dagger }a\rangle $. We observe that the qubit mean
excitation number $\langle \sigma _{q}^{+}\sigma _{q}^{-}\rangle $ and $%
G_{q}^{(2)}$ almost coincide at any time in the nondissipative case, as
shown in Fig.$\,$7(a). This is a signature of almost perfect two-qubit
correlation: if one qubit is excited, the other is also excited \cite{R9,R39}%
. This behavior indicates that the photon is directly absorbed by the two
qubits jointly. Fig.$\,$7(b) shows the influence of dissipation on the time
evolution of the equal-time two-qubit correlation function. Compared with the
mean qubit excitation number, the equal-time two-qubit correlation function
seemes to be more fragile to the system damping. However, during the time
evolution, it is notable that the two-qubit correlation function $G_{q}^{(2)}$
goes almost to zero when the cavity mean photon number $\langle a^{\dagger
}a\rangle $ is at maximum. This implies that the cavity photon is not absorbed
by a single qubit instead the two qubits jointly \cite{R40}.

\section{conclusion}

In summary, we have shown that a system consisting of a two-level atom and
a $\Lambda $-type or V-type three-level atom off-resonantly coupled to a cavity
mode can exhibit anomalous vacuum Rabi oscillations, indicating that two
atoms can be simultaneously excited by a single photon and jointly emit a
photon into the cavity mode in a reversible and coherent process.
Furthermore, we have shown a scheme to realize this interesting phenomena in
a circuit QED architecture, and study the quantum statistics of , in
particular, the time evolution of equal-time second-order correlation
function. Our numerical calculation shows a clear signature that a single
photon can be directly absorbed by two atoms simultaneously. In addition, we
show that multi-photons and even a classical field can also simultaneously
excite two atoms.

These process can be exploited for realizing efficient atom-atom or
atom-photon entanglement source \cite{R9}. It can also be used for the implementation
of novel schemes for the control and manipulation of atomic and cavity
states. The resonant coupling between one photon and two atoms may paves the
way for investigating multi-atom interaction mediated by cavity photon or
multi-qubit mixing process \cite{R57}. Although, the model we introduced in
this work is at single atom level, one can also apply this model to hybrid
quantum system formed by spin or atomic ensemble \cite{R58}. This opens up
a new possibility for quantum information on hybrid quantum system.

\acknowledgements This work was partly supported by the the NKRDP of China
(Grant No. 2016YFA0301802), NSFC (Grant No. 11504165, No. 11474152, No.
61521001).

\appendix
\section{}

In this appendix, we provide a detailed derivation of the effective
Hamiltonian in Eq.$\,$(4). We start from the original Hamiltonian in Eq.(1)
which is composed of an unperturbed part $H_{0}$ with known eigenvalues and
eigenstates and a small perturbation part $H_{I}$.  Our system operates
in the dispersive regime, where the atom-cavity detuning is larger than the
coupling strengths between them. We have $|\Delta _{p}|=|\omega _{i}-\omega
_{c}|\gg g_{p},$ $|\Delta _{s}|=|\omega _{i}-\omega _{e}-\omega _{c}|\gg
g_{s},$ and $|\Delta _{2}|=|\omega _{2}-\omega _{c}|\gg g_{2}$. The
frequency of the cavity mode satisfies the frequency matching condition $%
\omega _{c}\approx \omega _{e}+\omega _{2}$. For easy reference, we set $%
\omega _{g}=0$ hereafter.

Considering the system is operated in the dispersive regime, we can eliminate the
direct atom-cavity coupling using a unitary transformation
\begin{eqnarray}
\begin{aligned}
H_{eff}=e^{-X}He^{X},
\end{aligned}
\end{eqnarray}
where $X$ is chosen such that the direct coupling between the atom-cavity $%
H_{I}$ in the transformed Hamiltonian disappear. Here, we choose
\begin{eqnarray}
\begin{aligned}
X=\frac{g_{p}}{\Delta_{p}}a^{\dagger}\left|g\right\rangle\left\langle i\right| + \frac{g_{s}}{\Delta_{s}}a^{\dagger}\left|e\right\rangle\left\langle i\right| +\frac{g_{2}}{\Delta_{2}}a^{\dagger}\sigma_{2}^{-}-H.c.,
\end{aligned}
\end{eqnarray}
so that it satisfies $[H_{0},X]=-H_{I}$. By expanding to the third order of
the small parameters ($\frac{g_{p}}{\Delta _{p}},\frac{g_{s}}{\Delta _{s}},%
\frac{g_{2}}{\Delta _{2}}$), we have
\begin{eqnarray}
\begin{aligned}
&e^{-\lambda X}(H_{0}+\lambda H_{I})e^{\lambda X} =(H_{0}+\lambda H_{I})\\&+\lambda[(H_{0}+\lambda H_{I}),X]+\frac{\lambda^{2}}{2!}[[(H_{0}+\lambda H_{I}),X],X]
\\&+\frac{\lambda^{3}}{3!}[[[(H_{0}+\lambda H_{I}),X],X],X]+O(\lambda^{4})
\\&=H_{0}+\frac{\lambda^{2}}{2}[H_{I},X]+\frac{\lambda^{3}}{3}[[H_{I},X],X]+O(\lambda^{4})
,
\end{aligned}
\end{eqnarray}
where $\lambda $ is introduced to show the orders in the perturbation
expansion, and would be set to 1 after the calculations. Under the frequency
matching condition, by keeping resonant terms only (in the rotating wave
approximation), one can obtain the effective Hamiltonian:
\begin{eqnarray}
\begin{aligned}
H_{eff}=&\widetilde{\omega}_{c}a^{\dagger}a+\widetilde{\omega}_{1} \sigma_{1}^{+}\sigma_{1}^{-}+\widetilde{\omega}_{2} \sigma_{2}^{+}\sigma_{2}^{-}+\widetilde{\omega}_{i}|i\rangle_{1}\langle i|
\\&+( \chi a \sigma_{1}^{+} \sigma_{2}^{+} + H.c.),
\end{aligned}
\end{eqnarray}%
where
\begin{eqnarray}
\begin{aligned}
\chi=\frac{g_{s}g_{2}}{3}(\frac{1}{\Delta_{s}} + \frac{1}{\Delta_{2}})\frac{g_{p}}{\Delta_{p}} + \frac{g_{s}g_{p}}{3}(\frac{1}{\Delta_{s}} + \frac{1}{\Delta_{p}})\frac{g_{2}}{\Delta_{2}},
\end{aligned}
\end{eqnarray}
and $\sigma _{1}^{\pm }$ are the ladder operators for a reduced two-level
system formed by the lowest two levels ($|g\rangle _{1},|e\rangle _{1}$) of
the three-level atom. $\widetilde{\omega }_{c}=\omega _{c}-\frac{g_{p}^{2}}{%
\Delta _{p}}-\frac{g_{2}^{2}}{\Delta _{2}}$, $\widetilde{\omega }_{2}=\omega
_{2}+\frac{g^{2}}{\Delta _{2}}+\frac{2g^{2}}{\Delta _{2}}a^{\dagger }a$ ,$%
\widetilde{\omega }_{i}=\omega _{i}+(\frac{g_{p}^{2}}{\Delta _{p}}+\frac{%
g_{s}^{2}}{\Delta _{s}})+(\frac{g_{p}^{2}}{\Delta _{p}}+\frac{g_{s}^{2}}{%
\Delta _{s}})a^{\dagger }a$, and $\widetilde{\omega }_{1}=\omega _{e}+(\frac{%
g_{p}^{2}}{\Delta _{p}}-\frac{g_{s}^{2}}{\Delta _{s}})a^{\dagger }a$ are the
renormalization transition frequencies of the cavity and two atoms, which is
due to atom-cavity dispersive coupling. Since our system is initially
prepared in $|1,g,g\rangle$, it is apparently that $|i\rangle _{1}$ is
decoupled from our system, leading to a negligible occupation probability on
the this level in the whole process. Therefore, we can rewrite the
effective Hamiltonian in Eq.(A4)
\begin{eqnarray}
\begin{aligned}
H_{eff}=&\widetilde{\omega}_{c}a^{\dagger}a+\widetilde{\omega}_{1} \sigma_{1}^{+}\sigma_{1}^{-}+\widetilde{\omega}_{2} \sigma_{2}^{+}\sigma_{2}^{-}
\\&+( \chi a \sigma_{1}^{+} \sigma_{2}^{+} + H.c.).
\end{aligned}
\end{eqnarray}%

It is worth to point out when we transform Hamiltonian with a unitary
operator, the state (basis) also need to be transformed. This means that the operator in Eq.$\,$(A4) and Eq.$\,$(A6) are the dressed-state operators, i.e., operator represented via dressed-state basis (transformed basis). However, our system operates in the
dispersive regime, the dressed-state can be well approximated by
the bare states \cite{R59}, which are the eigenstates of the uncoupled Hamiltonian $%
H_{0}$.

\section{}

Using the similar procedures in section II, we can obtain the effective
Hamiltonian of a system consisting of a two-level atom and a V-type
three-level atom off-resonantly coupled to a cavity mode as depicted in the
Fig.$\,$1(b) of the main text. Similar to the previous procedure, we start
from the Hamiltonian ($\hbar =1$),
\begin{eqnarray}
\begin{aligned}
&H=H_{0}+H_{I}
\\&H_{0}=\omega_{c}a^{\dagger}a+\sum_{j=g,e,i}\omega_{j}|j\rangle_{1}\langle j|+\omega_{2}\sigma_{2}^{+}\sigma_{2}^{-},
\\&H_{I}=g_{p}a|e\rangle_{1}\langle i|+g_{s}a|g\rangle_{1}\langle i|+g_{2}a\sigma_{2}^{+}+H.c.
\end{aligned}
\end{eqnarray}
For easy reference, we define $\omega _{i}=0$ hereafter. When atom-cavity detuning are large enough, $|\Delta _{2}|=|\omega
_{2}-\omega _{c}|\gg g_{2},|\Delta _{p}|=|\omega _{e}-\omega _{c}|\gg g_{p},$%
and $|\Delta _{s}|=|\omega _{g}-\omega _{c}|\gg g_{s}$. We can eliminate the
direct atom-cavity coupling using the unitary transformation
\begin{eqnarray}
\begin{aligned}
&H_{eff}=e^{-S}He^{S},
\\&S=\frac{g_{p}}{\Delta_{p}}a^{\dagger}\left|i\right\rangle\left\langle e\right| + \frac{g_{s}}{\Delta_{s}}a^{\dagger}\left|i\right\rangle\left\langle g\right| +\frac{g_{2}}{\Delta_{2}}a^{\dagger}\sigma_{2}^{-}-H.c.,
\end{aligned}
\end{eqnarray}
where $S$ satisfies $[H_{0},S]=-H_{I}$. Further, we assume that the
frequency of the cavity mode is $\omega _{c}\approx \omega _{e}-\omega
_{g}+\omega _{2}$. Following the same procedure as that in Appendix A, we can obtain an effective
Hamiltonian

\begin{eqnarray}
\begin{aligned}
H_{eff}=&\widetilde{\omega}_{c}a^{\dagger}a+\widetilde{\omega}_{1}\sigma_{1}^{+}\sigma_{1}^{-}+\widetilde{\omega}_{2}\sigma_{2}^{+}\sigma_{2}^{-}+
\\&+(\chi a \sigma_{1}^{+}\sigma_{2}^{+} + h.c.),
\end{aligned}
\end{eqnarray}
where
\begin{eqnarray}
\begin{aligned}
\chi=\frac{g_{s}g_{p}}{3}(\frac{1}{\Delta_{p}} + \frac{1}{\Delta_{s}})\frac{g_{2}}{\Delta_{2}} + \frac{g_{s}g_{2}}{3}(\frac{1}{\Delta_{s}} + \frac{1}{\Delta_{2}})\frac{g_{p}}{\Delta_{p}}
\end{aligned}
\end{eqnarray}
is the effective coupling strength between the two atom and cavity, and $%
\sigma _{1}^{\pm }$ are the ladder operators for a reduced two-level system
formed by the two levels ($|g\rangle _{1},|e\rangle _{1}$) of the
three-level atom. $\widetilde{\omega}_{c}=\omega_{c}+\frac{g_{s}^{2}}{%
\Delta_{s}}-\frac{g_{2}^{2}}{\Delta_{2}}$, $\widetilde{\omega}%
_{2}=\omega_{2}+\frac{g_{2}^{2}}{\Delta_{2}}+\frac{2g_{2}^{2}}{\Delta_{2}}%
a^{\dagger}a$ and $\widetilde{\omega}_{1}=(\omega_{e}-\omega_{g})+(\frac{%
g_{p}^{2}}{\Delta_{p}}-\frac{g_{s}^{2}}{\Delta_{s}})+ (\frac{g_{p}^{2}}{%
\Delta_{p}}-\frac{g_{s}^{2}}{\Delta_{s}})a^{\dagger}a$ are the
renormalization transition frequencies of the cavity and two atoms.
The last term in Eq.$\,$(B3) implies that one can simultaneously excite two
atoms with one cavity photon in the system depicted in Fig.$\,$1(b).

\section{}

We study the influence of cavity decay and atom relaxation on the process by
solving the master equations. By including cavity decay and atom relaxation
terms we obtain the master equation:
\begin{eqnarray}
\begin{aligned} \frac{d\rho}{dt}=&-i[H,\rho]+\kappa
\mathcal{L}[a]+\sum_{q=1,2}(\gamma^{(q)}_{ge}\mathcal{L}[|g\rangle
_{q}\langle e|]\\&+\gamma^{(q)}_{ei}\mathcal{L}[|e\rangle _{q}\langle
i|]+\gamma^{(q)}_{gi}\mathcal{L}[|g\rangle _{q}\langle i|]),
\end{aligned}
\end{eqnarray}
where the Hamiltonian $H$ is given in Eq.$\,$(14), $\rho$ is the reduced density matrix of the system, $\mathcal{L}[O]=O\rho O^{\dagger }-O^{\dagger }O\rho/2-\rho O^{\dagger }O/2$, $\kappa $ and $%
\gamma _{jk}^{(q)}$ denote the photon decay rate and the relaxation rate of the $(|j\rangle _{q},|k\rangle _{q} )$ two level systems, respectively.

\end{document}